\RequirePackage{fix-cm}
\documentclass[twocolumn]{svjour3}         
\smartqed  % flush right qed marks, e.g. at end of proof
\usepackage{graphicx}
\usepackage{mathptmx}  
\usepackage{soul}
\usepackage{amsmath}
\usepackage{color}
\usepackage{soul}
\usepackage{latexsym}
\usepackage{braket}
\usepackage{hyperref}

\journalname{Granular Matter}
\begin{document}
%-|-%-|-%-|-%-|-%-|-%-|-%-|-%-|-%-|-%-|-%-|-%-|-%-|-%-|-%-|-%-|-%-|-%-|-%-|-%-|-%

\title{Jamming transition in non-spherical particle systems: pentagons vs. disks}
\author{Yiqiu Zhao         \and
        Jonathan Bar\'{e}s    \and
        Hu Zheng \and
        Cacey Stevens Bester  \and
        Yuanyuan Xu \and
        Joshua E. S. Socolar \and
        Robert P. Behringer
}
\institute{Yiqiu Zhao \and Jonathan Bar\'{e}s \and Hu Zheng \and Cacey Stevens Bester \and Yuanyuan Xu \and Joshua E. S. Socolar \and Robert P. Behringer
            \at   Department of Physics \& Center for Non-linear and Complex Systems, Duke University, Durham, NC, 27708, USA
            \and
            Jonathan Bar\'{e}s \at
            Laboratoire de M\'{e}canique et G\'{e}nie Civil, Universit\'{e} de Montpellier, CNRS, Montpellier, France 34090\\\email{jb@jonathan-bares.eu}
            \and
            Hu Zheng \at
            Department of Geotechnical Engineering, College of Civil Engineering, Tongji University, Shanghai, 200092, China\\\email{tjzhenghu@gmail.com}
            \and
            Cacey Stevens Bester \at
            Department of Physics and Astronomy, Swarthmore College, Swarthmore, PA 19081, USA 
            \and
            Yuanyuan Xu \at
            Department of Physics and Astronomy, Johns Hopkins University, 3400 N. Charles Street, Baltimore, MD 21218, US
}

\date{\today}
\maketitle
%-|-%-|-%-|-%-|-%-|-%-|-%-|-%-|-%-|-%-|-%-|-%-|-%-|-%-|-%-|-%-|-%-|-%-|-%-|-%-|-%

\begin{abstract}
We investigate the jamming transition in a quasi-2D granular material composed of regular pentagons or disks subjected to quasistatic uniaxial compression. We report six major findings based on experiments with monodisperse photoelastic particles with static friction coefficient $\mu\approx 1$. (1) For both pentagons and disks, the onset of rigidity occurs when the average coordination number of non-rattlers, $Z_{nr}$, reaches $3$, and the dependence of $Z_{nr}$ on the packing fraction $\phi$ changes again when $Z_{nr}$ reaches $4$. (2) Though the packing fractions $\phi_{c1}$ and $\phi_{c2}$ at these transitions differ from run to run, for both shapes the data from all runs with different initial configurations collapses when plotted as a function of the non-rattler fraction. (3) The averaged values of $\phi_{c1}$ and $\phi_{c2}$ for pentagons are around $1\%$ smaller than those for disks. (4) Both jammed pentagons and disks show Gamma distribution of the Voronoi cell area with same parameters. (5) The jammed pentagons have similar translational order for particle centers but slightly less orientational order for contacting pairs compared to jammed disks. (6) For jammed pentagons, the angle between edges at a face-to-vertex contact point shows a uniform distribution and the size of a cluster connected by face-to-face contacts shows a power-law distribution.
\end{abstract}


%-|-%-|-%-|-%-|-%-|-%-|-%-|-%-|-%-|-%-|-%-|-%-|-%-|-%-|-%-|-%-|-%-|-%-|-%-|-%-|-%

\section{Introduction}
The jamming transition for a granular material separates fluid-like states with zero yield stress from solid-like states that can support finite stress. The past two decades have seen a significant effort directed toward understanding the jamming transition in model granular systems consisting of spherical particles \cite{OHern2003_pre,Liu2010_arcmp,Silbert2010_sm,Bi2015_arcmp,Bi2011_nat,torquato2010_rmp}, due both to their relative simplicity and relevance for understanding glasses and suspensions \cite{parisi2010_rmp,OHern2003_pre,silbert2002_pre,Han2016_natcom,Peters2016_nat,sarkar2015_pre}. However, real world industrial and environmental processes usually involve particles that are not spherical, and recent work has shown that such particles can differ significantly in their geometrical and mechanical properties \cite{donev2004_nature,donev2007_pre,azema2007_pre,athanassiadis2014_soft,torquato2010_rmp,zhao2016_gm}. Ellipsoids typically jam at higher packing fraction than spheres \cite{donev2004_nature}, for example, and the modulus of a granular material can vary more than an order of magnitude when the particle shapes are changed while other properties are held fixed \cite{athanassiadis2014_soft}. Despite the increasing amount of attention given in recent years to the jamming of non-spherical particles \cite{donev2004_nature,donev2007_pre,vanderwerf2018_pre,smith2011_pre,zhang2013_acs,schaller2015_pre,saint2012_epl,torquato2010_pre,azema2012_pre,jaeger2015_soft,borzsonyi2013_soft,mailman2009_prl,azema2010_pre,jiao2009_pre,torquato2009_nature,han2015_epl}, significant questions remain. In particular, to our knowledge, few of the previous works regarding the jamming transition of non-spherical particles, especially with non-smooth boundary, have consisted in experimental studies \cite{tang2016_epl,zheng2017_epj,xu2017_epj}. We report here on experiments that provide relevant data on both macroscopic properties and microstructures of jammed packings of pentagonal particles. 

The pentagon is the regular polygon with smallest number of sides that cannot fill space, so it is often used as a model system for non-spherical particles with a non-smooth boundary \cite{azema2007_pre,estrada2011_pre,schilling2005_pre,duparcmeur1995_jpcm}. We report detailed comparisons between the properties of disks and regular pentagonal particles undergoing a jamming transition induced by uniaxial compression. We use photoelastic particles, which allow direct imaging of contacts and force network structures \cite{Daniels2017_rsi,Majmudar2007_prl,all2019_wiki}. We focus on two aspects of jamming: ($i$) the features of jamming transition, including the jamming packing fraction and the quasi-static evolution of non-affine displacements, contact numbers and  {contact} networks; and ($ii$) the packing structure after the jamming transition. In Section~2, we describe our experimental set-up. In Section~3, we report our observations and compare them with other works, and in Section~4, we summarize our findings.

%-|-%-|-%-|-%-|-%-|-%-|-%-|-%-|-%-|-%-|-%-|-%-|-%-|-%-|-%-|-%-|-%-|-%-|-%-|-%-|-%

\section{Experiment}

We perform experiments on two types of systems. The first consists of about $800$ identical regular pentagons and the second of about $900$ identical disks. The side length of the pentagons is $9.40\,$~mm while the radius of the disks is $6.35\,$~mm. The inter-particle friction coefficients for pentagons and disks are $\mu_{p}\approx 1.23$ (face to face) and $\mu_{d}\approx 1.07$ respectively. Both particle types are prisms of height $6.35\,$mm. The particles rest on a Plexiglas plate, and the particle-plate static friction coefficient is $\mu \approx 0.36$ for disks and $\mu \approx 0.59$ for pentagons. The particles are homemade, using the same method as in \cite{Cox2016_epl,bares2017_epj,all2019_wiki}. The Young's modulus of all particles is $2.9$~MPa.

\begin{figure}[!]
\centering \resizebox{0.95\hsize}{!}{\includegraphics{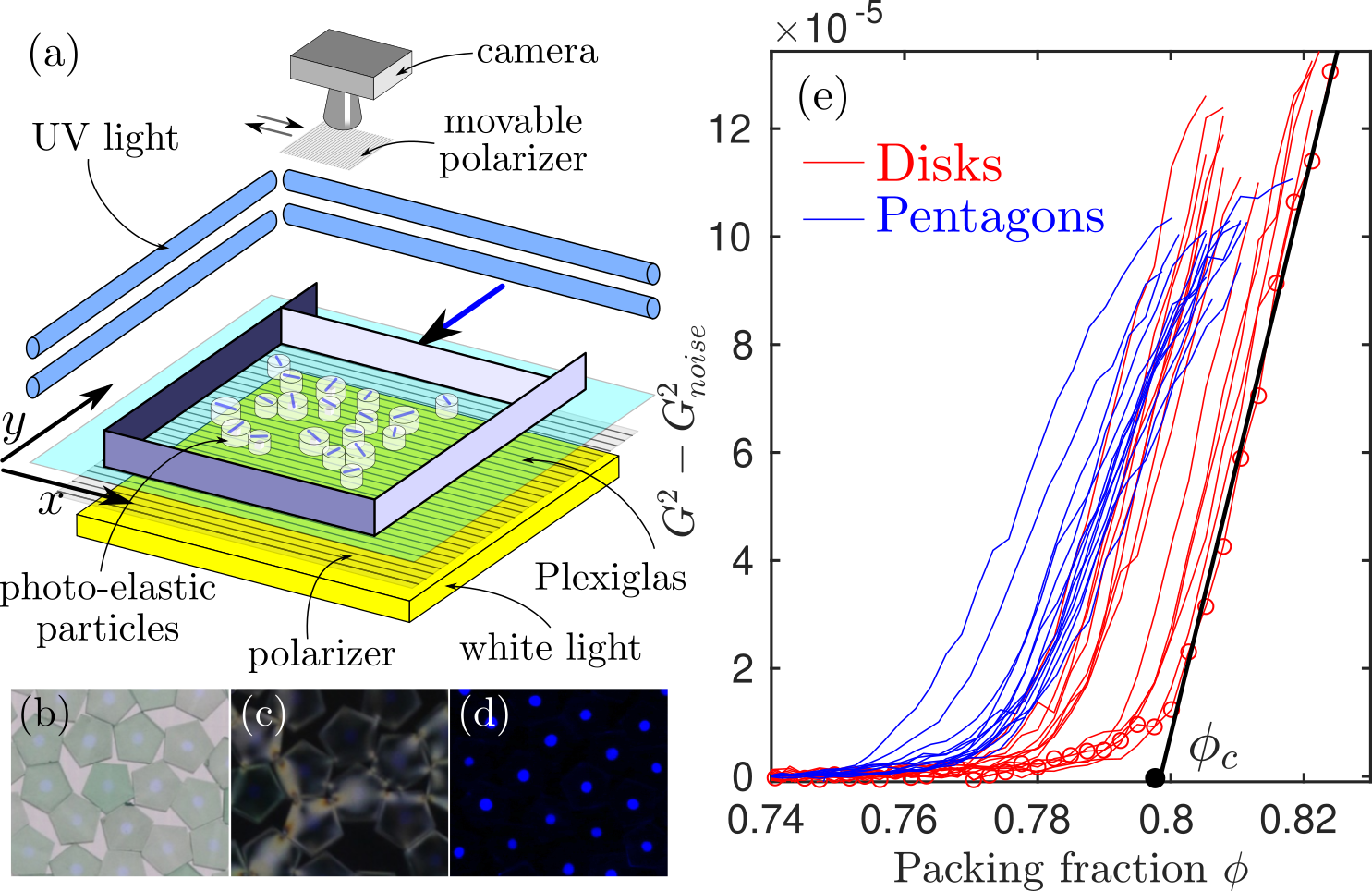}}
\caption{(a) The uniaxial compression experimental set-up and the imaging system. (b) Part of an example white light image. (c) Part of an example polarized light image. (d) Part of an example ultra-violet light image. (e) The pressure response of the granular system under compression measured by $G^2\equiv \braket{|\nabla I|^2}$, where $I$ is the light intensity of the polarized light image and $\braket{}$ means averaging over pixels \cite{Geng2001_prl}. Each curve corresponds to one uniaxial compression with a random initial configuration. Data from 14 runs of pentagons and 14 runs of disks are plotted. The black straight line shows a linear fit for the rightmost curve, where the data taken at each compression step are also shown. Its intersection with the noise level defines a reference packing fraction $\phi_c$ for that run.}
\label{Setup}
\end{figure}

The experimental set-up is similar to the one used in \cite{Cox2016_epl} as shown in Fig.~\ref{Setup}(a). For each run, particles are placed randomly into a rectangular box before compression to form a dilute unjammed state. The initial size of the box is $\sim 60$~cm $\times$ $\sim 40$~cm. Quasi-static uniaxial compression is achieved by moving one boundary inward in steps of $dl = 1$~mm, resulting in a $\delta\phi \sim 0.002$ change of packing fraction. After each compression step, three pictures are recorded by a digital camera (Canon EOS 70D with $5472 \times 3648$~px$^2$) under different lighting conditions: ($i$) white light image used to find the particle boundaries (see Fig.~\ref{Setup}(b)); ($ii$) polarized light image used to find the stress and contacts (see Fig.~\ref{Setup}(c)) and ($iii$) ultra-violet (UV) light image, used to find the particle centers, which have been marked with UV sensitive labels (see Fig.~\ref{Setup}(d)). For both particle shapes, we perform $14$ different compression experiments starting from different initial random stress free configurations.

In this work, we determine the pressure in the system by measuring $G^2\equiv \braket{|\nabla I|^2}$ \cite{Geng2001_prl,Daniels2017_rsi,all2019_wiki,zhao2019_npj}, where $I$ is the light intensity of the green channel of the polarized image. The $\braket{}$ average is taken over the pixels within the particles. $G^2$ has been shown to be proportional to the pressure for both disks and pentagons  {under uniaxial compression}~\cite{Geng2001_prl}. In addition, we collect statistics on the numbers and types of contacts between particles. Two particles are determined to be in contact if the distance between them is smaller than a threshold value and, on both sides of the contact, the intensity of the polarized image is larger than a threshold \cite{Majmudar2007_prl,Daniels2017_rsi}. This threshold is tuned carefully to be small enough so that all stressed contacts identifiable by the human eye are detected by the algorithm. 

We also consider the deviations of particle displacements from their expected motions if the compression were to induce only affine distortions of the packing. We measure \linebreak$\delta x_{rms}(\phi)~\equiv~[\overline{dx(\phi)^2}]^{1/2}$, where $dx(\phi)$ is the $x$-component (transverse to the compression direction) of the displacement of a given particle during the compression step $(\phi,\phi+\delta\phi)$. The bar indicates an average over all particles.

%-|-%-|-%-|-%-|-%-|-%-|-%-|-%-|-%-|-%-|-%-|-%-|-%-|-%-|-%-|-%-|-%-|-%-|-%-|-%-|-%

\section{Results}

%-|-%-|-%-|-%-|-%-|-%-|-%
\subsection{Jamming transition}

\begin{figure}[!]
\centering \resizebox{0.95\hsize}{!}{\includegraphics{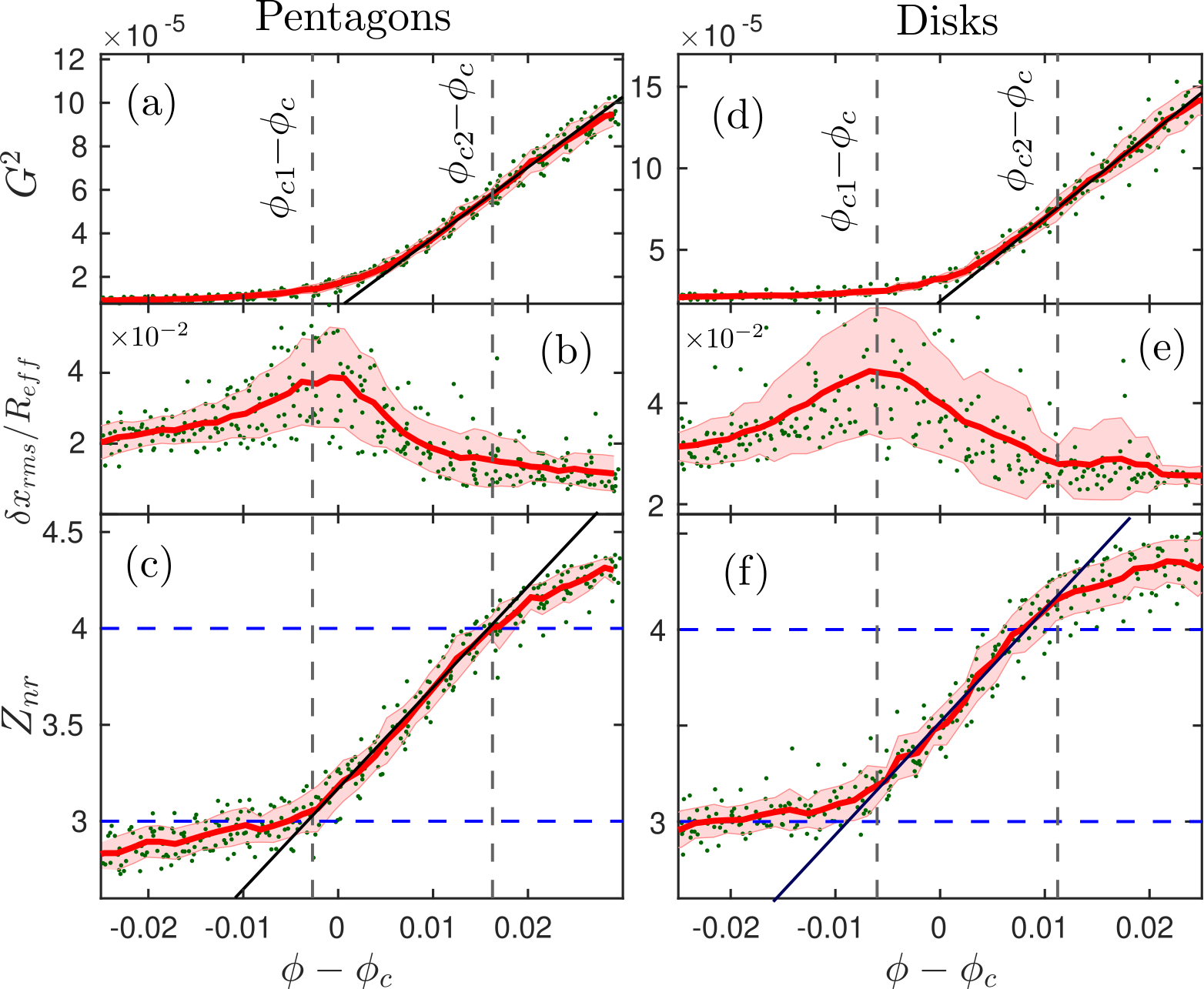}}
\caption{In all sub-figures, the green dots represent the original data with different initial conditions, the red curves represent the averaged data and the shaded areas show standard deviations. (a-c) show pentagon measurements and (d-f) show disk measurements. (a) and (d) show the evolution of $G^2$ (proportional to pressure \cite{Geng2001_prl}) near the jamming transition. The black lines are linear fits for the averaged data with $\phi>\phi_{c2}$. (b) and (e) show the evolution of the root mean square non-affine displacement $\delta x_{rms}$ at step $(\phi,\phi+\delta\phi)$. (c) and (f) show the evolution of the mean non-rattler contact number $Z_{nr}$. The black lines are linear fits for data with $\phi_{c1}<\phi<\phi_{c2}$. The blue dashed lines show where $Z_{nr}=3$ and $Z_{nr}=4$.}
\label{contact}
\end{figure}

The jamming transition happens when a granular system first supports non-zero stress as the compression proceeds. Fig.~\ref{Setup}(e) shows the dependence of the pressure on the packing fraction $\phi$. Consistent with the results reported in \cite{bandi2013_pre}, for each run of experiment, $G^2(\phi)$ increases smoothly to reach a clearly defined linear regime. For each run, we infer a reference packing fraction $\phi_c$ corresponding to the packing where a fit to the linear regime of $G^2(\phi)$ reaches the noise level, as demonstrated in Fig.~\ref{Setup}(e). The mean value of $\phi_c$ for pentagons was found to be $0.771 \pm 0.005$ and for disks was $0.788\pm0.007$.

Figures~\ref{contact}(a) and~\ref{contact}(d) show $G^2$ as a function of $\Delta\phi = \phi-\phi_c$ for pentagons and disks, respectively. In both cases the data from many runs collapse to a single curve with similar features for the two particle types. The green dots show data from individual runs, and the red lines are averages over all runs at the same $\Delta\phi$. The black curves show the linear fits expected from Fig.~\ref{Setup}(e). As shown in Fig.~\ref{Setup}(e), $G^2$ begins to grow before $\phi_c$ is reached, making it difficult to use $G^2$ to identify the jamming transition. The effect arises because as the wall moves, a growing  layer of jammed particles is pushed along the moving wall until it reaches the opposite wall, at which point the system actually jams globally. The non-zero $G^2$ before jamming comes from the weak force network required to move this jammed layer given the small basal friction force on it.

We determine the jamming packing fraction by measuring  {$\phi$ at} the transition signaled by  {the change in behavior of} the mean contact number of non-rattlers $Z_{nr}$,  {where ``non-rattlers'' are defined here as all particles that have detectable contacts with at least two neighboring particles.}  {The }dependence  {of $Z_{nr}$} on $\phi-\phi_c$ is plotted in Figs.~\ref{contact}(c) and (f) for pentagons and disks respectively. We observe transitions near $Z_{nr}=3$ and  $Z_{nr}=4$, corresponding to packing fractions at $\phi_{c1}$ and $\phi_{c2}$, respectively. A linear regime of $Z_{nr}(\phi)$ is observed between the two transitions, shown by the black fit lines. We define $\phi_{c1}$ and $\phi_{c2}$ as the packing fractions where this linear regime begins and ends. We call $\phi_{c1}$ the jamming packing fraction since it corresponds to the first fast rising regime of $Z_{nr}$. This corresponds to the point where the jammed layer hits the opposite wall, so that further compression induces rearrangements of the particles. Beyond $\phi_{c2}$, the  rate of increase of $Z_{nr}$ is smaller than in the linear regime. $\phi_{c1}$ and $\phi_{c2}$ also mark changes in the evolution of $\delta x_{rms}$, which reaches a peak at $\phi_{c1}$ and drops to a plateau for $\phi>\phi_{c2}$, as shown in Figs.~\ref{contact}(b) and (e). We refer to $\phi_{c2}$ as the ``stable'' packing fraction because beyond this point $\delta x_{rms}(\phi)$ takes its minimal value, showing that there are no further significant particle rearrangements. 

The existence of $\phi_{c1}$ and $\phi_{c2}$ has been observed in the uniaxial compression of bidisperse disks \cite{bandi2013_pre}. Here we provide contact number information at those transitions. Fig.~\ref{contact}(c) and (f) show that the average values of $\phi_{c1}$ and $\phi_{c2}$ are around $0.769$ and $0.788$ for pentagons and $0.781$ and $0.799$ for disks. The uncertainty of those estimates is around $0.002$. We note that both $\phi_{c1}$ and $\phi_{c2}$ are about $1$\% smaller for pentagons, which is consistent with recent numerical simulations \cite{wang2015_pre}.

A striking observation from Figs.~\ref{contact}(c) and~\ref{contact}(f) is that both pentagons and disks show $Z_{nr}(\phi_{c1})\approx3=d+1$ and $Z_{nr}(\phi_{c2})\approx4=2d$, where $d=2$ is the dimension of the packing.  {For disks, these two values correspond to saturated bounds on the numbers of constraints required to avoid floppy modes and assure geometric consistency, respectively.  For pentagons, however, the situation is more complicated, and it appears that the bounds are not saturated, as suggested by the following arguments.}  

 {Assume that the forces at each contact do not lie precisely at the Coulomb threshold (i.e., that an infinitesimal increase of the tangential component of contact force will not cause sliding).  For frictional disks, $Z_{nr}\geq d+1$ is the condition for the granular system to have no floppy mode~\cite{Hecke2009_jpcm}.} The fact that  {our disks}  jam close to $d+1$ indicates that the number of  contacts  {at the Coulomb threshold} is small, consistent with numerical simulations  {on particles with friction coefficients similar to ours}~\cite{Silbert2010_sm}.  {For frictional pentagons, the contact number required to eliminate all floppy modes depends on the number fraction, $n$, of face-to-face (ff) contacts. A ff contact constrains 3 degree of freedom (DOFs) that contribute to the energy of a configuration: relative translations both along and transverse to the shared face, and relative rotation. A face-to-vertex (fv) contact, on the other hand, constrains only the two relative translations.  The total number of constraints on the motions of $N$ particles is 
\begin{equation}
N_c = \bigg(3n+2(1-n)\bigg)\frac{NZ_{nr}}{2} \quad =~\frac{n+2}{2}NZ_{nr}.
\end{equation}
The total number of DOFs is $3N$, and to eliminate all floppy modes, $N_c$ must be greater than or equal to this.  Thus the isostatic condition becomes $Z_{nr}\geq 6/(n+2)$.  At $\phi_{c1}$, we find the averaged $n= 0.31\pm 0.02$, where the errorbar is the standard deviation. Therefore the isostatic value of $Z_{nr}$ becomes $6/(n+2)\approx 2.6$, which is significantly smaller than our observed value near 3.  We propose three possible reasons for this mismatch: (1) some contacts are at the Coulomb threshold, as observed in numerical simulations for disks \cite{henkes2010_epl}; (2) not all of the constraints are linearly independent; and (3) many of the constraints are one-sided; i.e., the normal forces support compression but not tension, which has implications for rotations as well as translations.
All three proposed reasons suggest that the naive constraint counting argument over-counts the number of constraints and thus underestimates of the contact number of an isostatic packing.}

\begin{figure}[!]
\centering \resizebox{0.99\hsize}{!}{\includegraphics{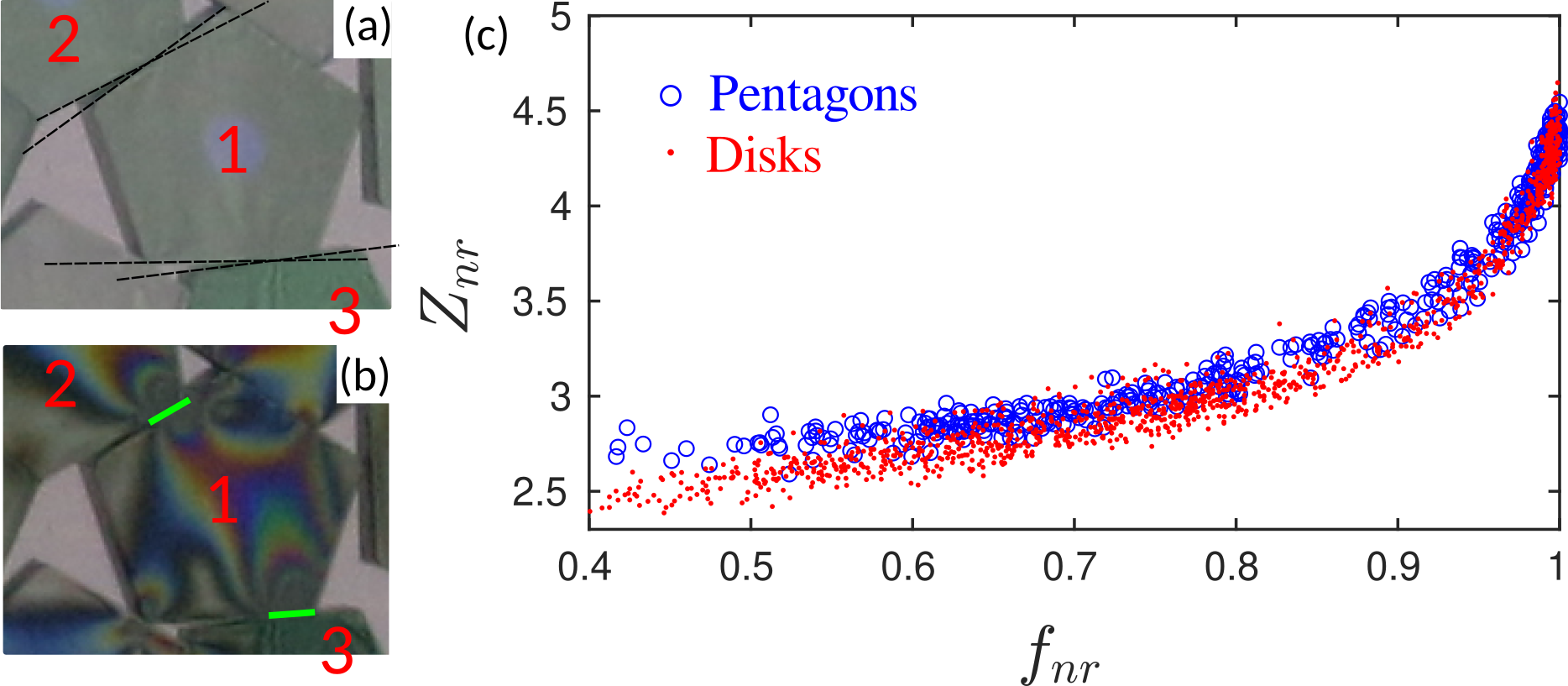}}
\caption{ {(a-b) Effect of particle deformability. (a) The contact between particle 1 and 2 (or 1 and 3) is a face to vertex contact. The black dashed lines show the orientations of the contacting sides. (b) An image showing the photoelastic fringes associated with the contacts. Small deformations near the contacts create finite contact areas (green bars).}~(c) The dependence of $Z_{nr}$ on the non-rattler fraction $f_{nr}$ for pentagons (blue circles) and disks (red dots) for all runs of experiments starting with different initial configurations.}
\label{fnr}
\end{figure}

 {The change in behavior at $\phi_{c2}$ is expected to be associated with the purely geometric constraint that disallows packings with overlapping particles.}  For rigid disks, the non-overlapping condition requires $Z_{nr}\leq 2d$ \cite{Hecke2009_jpcm,roux2000_pre}. Achieving higher values of $Z_{nr}$ requires substantial deformation of the particles. Thus, a change in behavior at $Z_{nr}=4$ may be expected for disks. For generic, non-spherical particles, the analogous upper bound is $Z_{nr}\leq d(d+1)=6$ \cite{Hecke2009_jpcm,roux2000_pre}.
 {It has been noted, however, that the bound given by the non-overlapping condition should depend on $n$ because one ff contact constrains two DOFs where one fv  contact constrains only one DOF \cite{azema2013_pre,marschall2018_pre,vanderwerf2018_pre}. Note that} 
 {These numbers differ from the above counting}
 {because virtual sliding does not cause overlap but does cost energy~\cite{henkes2010_epl}. The non-overlapping condition thus implies $Z_{nr}\leq 6/(1+n)$. At $\phi_{c2}$, we find $n=0.34\pm 0.02$, giving an upper bound on $Z_{nr}$ of approximately $4.5$. This is consistent with our observation that $Z_{nr}(\phi_{c2})\approx 4$, but raises the question of why the upper bound is not saturated.}  {We note that the above constraint counting arguments assume perfectly rigid particles, while our particles do deform slightly  {(with strain $< 5\%$)} under the forces achieved in our experiments.  {For example, fv contacts under large force can have finite contact area, as shown in Fig.~\ref{fnr}(a) and (b), which may be relevant for constraint counting.}  A theory that takes particle deformability is beyond the scope of this paper.}

The collapse of data in Fig.~\ref{contact} is achieved by shifting each run of data with respect to the reference point $\phi_c$. We find that the data also collapses when we plot the various quantities against the fraction of non-rattler particles, $f_{nr}$. It is defined as the fraction of particles having at least $2$ contacts. For example, Fig.~\ref{fnr}(c) shows $Z_{nr}(f_{nr})$ for pentagons and disks with all runs plotted. The runs with pentagons and disks collapse on two similar curves. We note that when $Z_{nr}\approx 3$, $f_{nr}\approx 0.8$, indicating that about $20$\% of the particles are not part of the jamming network, whereas when $Z_{nr}\approx 4$, almost all particles are non-rattlers. This provides a qualitative explanation for why large non-affine deformations occur near $\phi_{c1}$ and not above $\phi_{c2}$. The data collapse with $f_{nr}$ for different initial conditions in a system of disks was reported in \cite{Bi2011_nat}. Our results show that it holds for pentagonal particles as well.

%-|-%-|-%-|-%-|-%-|-%-|-%
\subsection{Jammed structure}

To further characterize the jammed packings, we collect data on spatial correlations of particle positions, bond orientations and the local environments of individual particles at the packing fraction $\phi_{c2}$. We found that any packing fraction above $\phi_{c1}$ yields similar results. 

\begin{figure}[!]
\centering \resizebox{0.9\hsize}{!}{\includegraphics{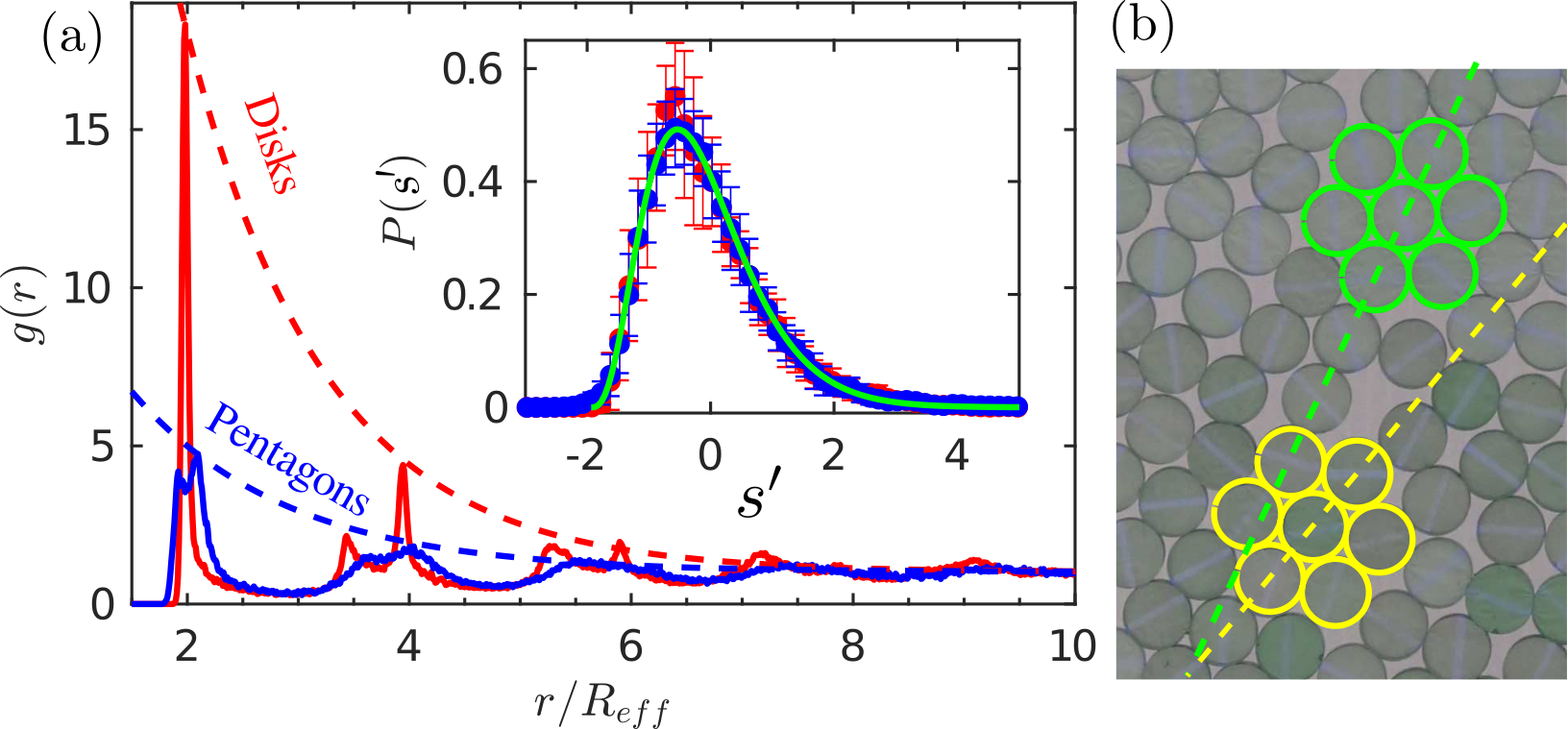}}
\caption{(a) Pair correlation functions $g(r)$ (solid curves) and their exponential envelope fitted to the first several peaks, for pentagons (blue curves) and discs (red curves). Insert: Probability density function of the rescaled Voronoi cell area $s'\equiv (s-\bar{s})/\sigma_s$, where $s$ is the Voronoi area and  $\bar{s}$  and $\sigma_s$ are its mean and standard deviation. The green solid curve is a Gamma distribution fit. Data are taken at $\phi_{c2}$. (b) Part of an example jammed disk packing, where two areas of strong local hexagonal order are highlighted using different color. Their misalignment is highlighted by the dashed lines.}
\label{pair}
\end{figure}

Figure~\ref{pair}(a) shows the pair correlation function $g(r)$ for pentagons and disks with the distances between particle centers scaled by an effective particle radius $R_{eff}$.  {When calculating $g(r)$, the particles with centers within a distance $r$ of the boundaries are excluded from the average.} For disks $R_{eff}$ is simply the particle radius, while for pentagons it corresponds to the radius of a disk of the same area. The dashed curves show an envelope of the form $g(r_{peak}) = a*e^{-r/l_c}+1$ for the heights of the first several peaks. For disks, we find $l_{c,d}/R_{eff}=1.24\pm 0.21$ and for pentagons, $l_{c,p}/R_{eff}=1.39\pm 0.5$. The values of $l_{c}/R_{eff}$ are not significantly different, indicating that the degrees of translational order in the two systems are roughly the same. 

\begin{figure}[!]
\centering \resizebox{1\hsize}{!}{\includegraphics{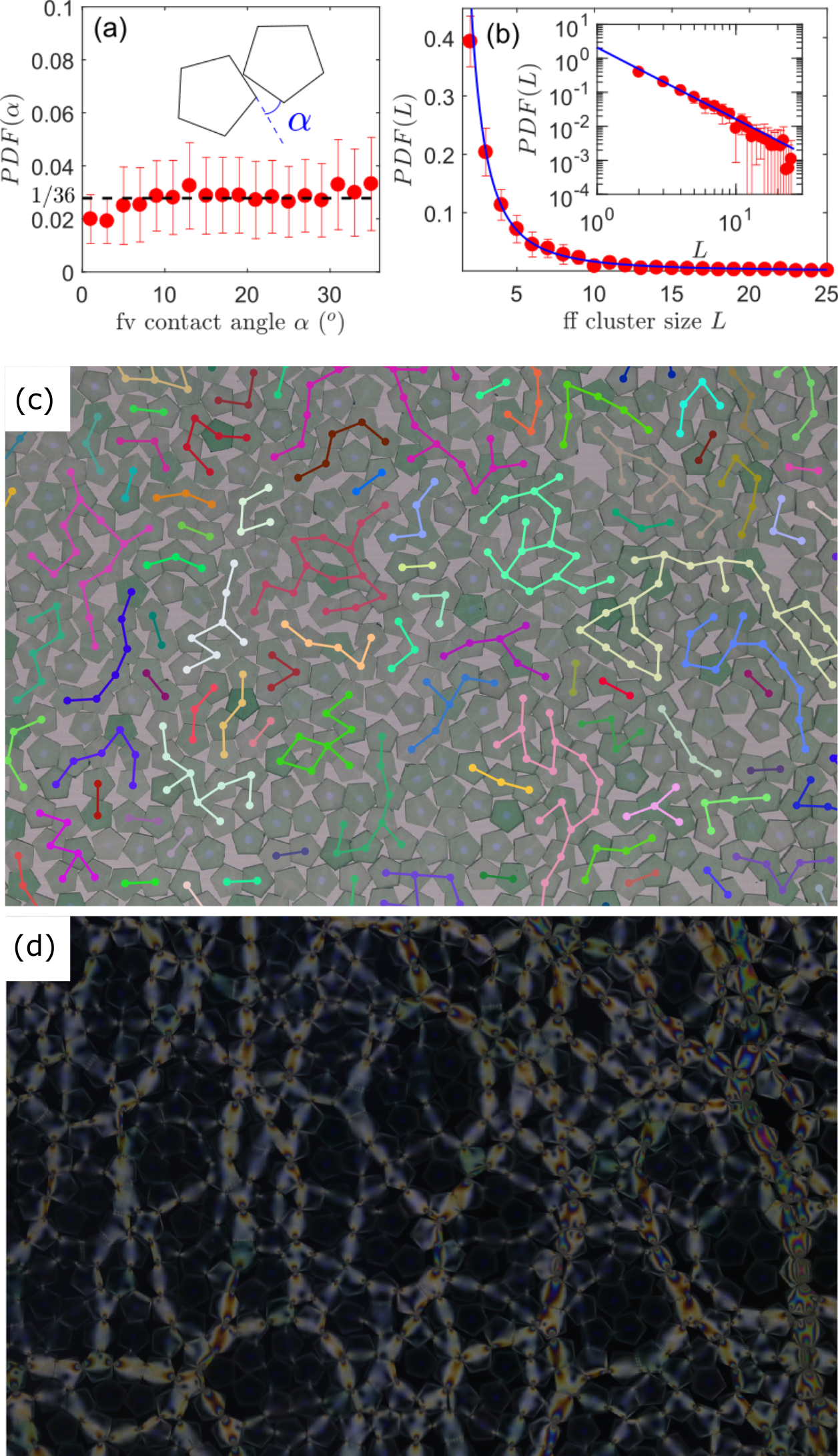}}
\caption{(a) The probability density function of face-to-vertex (fv) contact angle $PDF(\alpha)$, defined as the smallest angle formed at the fv contact as shown in the insert. The black dashed line shows a uniform distribution for comparison. (b) Distribution of the size of face-to-face (ff) connected particle clusters $PDF(L)$. $L$ is the number of particles in a ff cluster. Insert: Log-log plot of the data. In both cases, blue curves are power-law fits to the data with exponent $\beta=2.1\pm 0.04$. Data in both (a) and (b) are for packing fraction $\phi_{c2}$.  {(c) An example pentagon packing at $\phi_{c2}$. Particles that belong to same ff cluster are linked by line segments connecting their centers. Different clusters are labeled by different color chosen randomly. (d) Same packing in (c) viewed under polarizer, showing the stress distribution of the system. Brighter regions have higher stress and darker regions have smaller stress.}}
\label{angle}
\end{figure}

 In Fig.~\ref{pair}(a), $g(r)$ for jammed pentagons clearly shows a split first peak as numerically observed by \cite{nguyen2014_pre,wang2015_pre}. This may stem from the structure of the densest pentagon packing, in which each particle has six neighbors, including four face-to-face contacts and two contacts with $\alpha=\pi/5$ (see Refs.~\cite{duparcmeur1995_jpcm,wang2015_pre}).  {The definition of $\alpha$ is shown in the inset to Fig.~\ref{angle}(a).} This suggests that we compare the neighborhoods of jammed pentagons and disks at $\phi_{c2}$ by calculating the local ($l$) and global ($g$) $6$-fold bond orientational order parameter \cite{torquato2010_rmp,wang2015_pre}:
 
\begin{equation}
Q_{6}^{l}\equiv \frac{1}{N}\sum_{k=1}^N\left|\frac{1}{z_{k}}\sum_{j}^{z_{k}}e^{i6\theta_{kj}}\right|,~Q_6^{g}\equiv \left|\frac{1}{N_b}\sum_{bond}e^{i6\theta_{kj}}\right|,
\end{equation} 

\noindent where $\theta_{kj}$ is the angle of the vector joining the centers of particles $j$ and $k$ with respect to an arbitrary reference direction, $\sum_j^{z_{k}}$ means the sum over all particles $j$ that are in contact with particle $k$, $N$ is the total number of particles, $\sum_{bond}$ means the sum over all $kj$ pairs that are in contact and $N_b$ is the number of such bonds. We find for pentagons $Q_{6,p}^l=0.64 \pm 0.02$ and $Q_{6,p}^{g}=0.11 \pm 0.07$, and for disks $Q_{6,d}^l = 0.72 \pm 0.03$ and $Q_{6,d}^{g}=0.24 \pm 0.09$. The large values of $Q^{l}$ indicate that the centers of nearest neighbors of a particle tend to form hexagons. However, those hexagons tend not to be aligned with each other over larger distances (see Fig.~\ref{pair}(b)), resulting in small $Q^g$ values. We also note both $Q_6^l$ and $Q_6^g$ are slightly smaller for jammed pentagons, which is consistent with recent numerical simulations \cite{wang2015_pre}.  {
%We found no global orientational order for particle orientations for pentagons. 
We also check the global 5-fold bond orientational order parameter for pentagons $Q_{5,p}^g\equiv |\frac{1}{N_b}\sum_{bond}\exp(i5\theta_{kj})|=0.007\pm 0.003$ as well as the orientational order parameter for particle orientations $S_5\equiv |\frac{1}{N}\sum_{k=1}^N\exp(i5\beta_{k})|=0.02\pm 0.01$, where $\beta_k$ is the angle between the vector pointing from the center to one of the vertex of $k$th pentagon and the vertical direction.  We note both $Q_{5,p}^g$ and $S_5$ are close to zero, showing that there is no global pentagonal order for both bond and particle orientations.}

To study the statistics of local environments we measure the distribution of Voronoi cell area $s$ for both shapes. For a disk, the cell is defined as the associated Voronoi cell obtained from the set of points marking the centers of the disks. For the pentagon packing, we generalize the concept of Voronoi tesselation using the the definition detailed in \cite{wang2015_pre}. With this definition, the Voronoi cell for a pentagon is not necessarily a convex polygon. The insert of Fig.~\ref{pair}(a) shows the distribution of the rescaled cell area $s'\equiv (s-\bar{s})/\sigma_s$ for both pentagons (blue) and disks (red), with $\bar{s}$ and $\sigma_s$ being the average and standard deviation of $s$ in each case. Surprisingly, the two distributions collapse to a single curve. The green curve shows a gamma distribution fit to pentagon data (with same form as in \cite{cheng2010_soft}):

\begin{equation}
P(s')=\frac{1}{(s'-\bar{s}_0)\Gamma(k)}\left(\frac{(s'-\bar{s}_0)}{\theta}\right)^k \exp\left(\frac{-(s'-\bar{s}_0)}{\theta}\right),
\end{equation}

\noindent where the three fitting parameters are $\bar{s}_0=-1.94\pm 0.01$, $k=4.18\pm 0.08$ and $\theta = 0.442\pm 0.005$. The same kind of distribution has been reported in other stable granular systems with different $\phi$, including 3D monodisperse spheres \cite{aste2007_epl,aste2008_pre} and 2D poly-disperse tapioca pearls \cite{cheng2010_soft}. We show that this distribution also works for pentagons.

The rotational degree of freedom for pentagons gives rise to geometric features in pentagon packings that have no analogue in disk packings. For example, a description of the contact between two particles requires specification of an angle $\alpha$ defined as the smaller angle between the pentagon edges (see the insert in Fig.~\ref{angle}(a)). For face-to-vertex (fv) contacts (\textit{i.e.}, non-zero $\alpha$), we find that the distribution $PDF(\alpha)$ is uniform within the error bars. (See Fig.~\ref{angle}(a)). When $\alpha=0$ we have a face-to-face (ff) contacts. Any ff contacts that were misclassified by the automated $\alpha$ detection algorithm were identified by eye. However, for very small $\alpha$, one cannot distinguish between ff and fv contacts, which may explain why $PDF(\alpha)$ is slightly smaller for $\alpha$ close to zero. (See Fig.~\ref{angle}(a)). At $\phi_{c2}$, we find that $34\%\pm 2\%$ of the contacts are ff contacts. The presence of ff contacts raises the possibility of the propagation of orientational order through clusters of particles joined by such contacts. Figure~\ref{angle}(b) shows the distribution of the number of particles $L$ in a ff-connected cluster. The distribution follows a power-law, as indicated by the blue line representing a fit $P(L)\propto L^{-\beta}$ with $\beta=2.1\pm 0.04$. At $\phi_{c2}$, the average largest ff cluster size over different experiments is $34 \pm 13$ grains, which is only about 5\% of the number of particles in the system.  {Figure~\ref{angle}(c) plots the spatial distribution of the ff clusters for an example packing at $\phi_{c2}$. For completeness the polarized image of the same packing is also shown in Fig.~\ref{angle}(d), indicating the stress distribution of the system.}

%-|-%-|-%-|-%-|-%-|-%-|-%-|-%-|-%-|-%-|-%-|-%-|-%-|-%-|-%-|-%-|-%-|-%-|-%-|-%-|-%

\section{Conclusion}

We studied the jamming transition induced by quasi-static uniaxial compression of monodisperse granular systems consisting of regular pentagons or disks with static friction coefficient $\mu \approx 1$. We focused on two types of measurements: the evolution of state variables through the jamming process and the geometric structure of the jammed packings. The differences between the two particle shapes provide a means of identifying the properties of frictional granular materials that may be generic, along with those that are special to disks.

The quasi-static evolution of pressure, contact numbers, and non-affine displacement per step during uniaxial compression are qualitatively the same for pentagons and disks, despite the fact that the mean packing fractions $\phi_{c1}$, above which the system jams, and $\phi_{c2}$, above which the particles hardly ever rearrange, are $1$\% smaller for pentagons. We note two important findings: ($i$) We found $Z_{nr}(\phi_{c1})\approx3$ and $Z_{nr}(\phi_{c2})\approx4$ for both pentagons and disks.  {For disks, those two values are the lower (isostatic) and upper (non-overlapping) bounds of $Z_{nr}$ for jammed packings.}
 {For pentagons, however, the values of $Z_{nr}$ do not match  predicted bounds based on a simple constraint counting argument.}~($ii$) We found that all $Z_{nr}$ data collapse naturally when plotted against $f_{nr}$ for different initial configurations without any rescaling and the collapsed curves for both shapes are similar, consistent with the observation in \cite{Bi2011_nat}.

The geometric structures of the jammed packings of pentagons and disks were compared in detail, leading to three major observations: ($i$) The pair correlation functions for both shapes show the same characteristic decay length for the peak heights, indicating similar degrees of translational order. ($ii$) The 6-fold bond orientational order is smaller for jammed pentagons both locally and globally, consistent with recent numerical observations \cite{wang2015_pre}. ($iii$) The Voronoi cell area for both shapes follow the same Gamma distribution seen in other stable granular packings \cite{aste2007_epl,aste2008_pre,cheng2010_soft}.

The jammed pentagon packings have structural features with no analogue in disk packings, including the distributions of fv contact angles, $PDF(\alpha)$, and of the sizes of the ff clusters, $PDF(L)$. $PDF(\alpha)$ is constant and $PDF(L)$ shows a power-law distribution. The wide distribution $PDF(\alpha)$ reflects a complicated local packing structure, which tends to reduce the bond orientational order for pentagons as compared to disks. The rapid decay of $PDF(L)$ indicates that ff clusters are usually very small and do not percolate, consistent with recent numerical simulations \cite{nguyen2014_pre}.

The pentagonal particles used for this work have features, such as vertex-to-face or face-to-face contacts and complex geometrical packing constraints, that may be expected to occur in packings of real sands or rocks. The similarity of the results to those for frictional disks suggests that lessons learned from the simpler model systems are indeed useful conceptual guides to the analysis of real systems of grains with aspect ratios near unity.

\noindent\textbf{Acknowledgement} 

This work is dedicated to Bob Behringer, whom we are deeply indebted to and will forever miss. Though the paper was written after his passing, his role in supporting and mentoring this research justifies including him as a coauthor. Discussions with Yuchen Zhao, \'Emilien Az\'ema, Dong Wang, Ryan Kozlowski and Aghil Abed Zadeh are highly appreciated. This work was funded by NSFC Grant No. 4167\linebreak2256 (HZ), NSF Grant No. DMR1206351 and DMR1809762, ARO No. W911NF-18-1-0184, NASA Grant No. NNX15AD\linebreak38G, DARPA Grant No. 4-34728, the William M. Keck Foundation, and a Duke University Provost's Postdoctoral fellowship (CSB).

\bibliographystyle{unsrt}
\bibliography{reference}

\end{document}